\documentclass[naturefem]{nature}
\usepackage[T1]{fontenc}
\usepackage{graphicx}
\usepackage{color}
\usepackage[table]{xcolor}
\usepackage{tabularx}
\usepackage{subfigure}
\usepackage{bm}
\usepackage{soul}

\usepackage{lineno}
\usepackage{hyperref}

\usepackage[normalem]{ulem}
\useunder{\uline}{\ul}{}

\usepackage{caption}
\DeclareCaptionLabelSeparator{pipe}{ | }
\makeatletter
\let\saved@includegraphics\includegraphics
\AtBeginDocument{\let\includegraphics\saved@includegraphics}
\renewenvironment*{figure}{\@float{figure}}{\end@float}
\makeatother

\makeatletter
\let\saved@includegraphics\includegraphics
\AtBeginDocument{\let\includegraphics\saved@includegraphics}
\renewenvironment*{table}{\@float{table}}{\end@float}
\makeatother

\usepackage{array}
\newcommand{\PreserveBackslash}[1]{\let\temp=\\#1\let\\=\temp}
\newcolumntype{C}[1]{>{\PreserveBackslash\centering}p{#1}}
\newcolumntype{R}[1]{>{\PreserveBackslash\raggedleft}p{#1}}
\newcolumntype{L}[1]{>{\PreserveBackslash\raggedright}p{#1}}
      
\hbadness=\maxdimen
\vbadness=\maxdimen
\showboxdepth=\maxdimen
\showboxbreadth=\maxdimen
  
\newcommand{\comment}[1]{}
 
\begin{document}


\title{~~~~~GdBO$\bm{_{3}}$ and YBO$\bm{_{3}}$ Nanocrystals under Compression}
\maketitle

\begin{center}

\noindent 	Robin Turnbull$^{a,*}$,
			Daniel Errandonea$^{a}$,
			Juan \'{A}ngel Sans$^{b}$,
			Vanesa Paula Cuenca-Gotor$^{b}$,
			Rosario Isabel Vilaplana$^{c}$,
			Jordi Ib\'{a}\~nez$^{d}$,
			Catalin Popescu$^{e}$,	
			Agata Szczeszak$^{f}$,
			Stefan Lis$^{f}$ and
			Francisco Javier Manj\'{o}n$^{b}$



$^a$  	Departamento de F\'{i}sica Aplicada - Instituto de Ciencia de Materiales,
		MALTA Consolider Team,
		Universidad de Valencia, 
		Edificio de Investigaci\'{o}n, 
		C/Dr. Moliner 50, 
		Burjassot, 
		46100 Valencia, 
		Spain.
		\vspace{1mm}
\\ 
$^b$  	Instituto de Dise\~{n}o para la Fabricaci\'{o}n y Producci\'{o}n Automatizada,
		MALTA Consolider Team,
		Universitat Polit\`{e}cnica de Val\`{e}ncia,
		Cam\'{i} de Vera s/n, 
		46022 Valencia,
		Spain.
		\vspace{1mm}
\\
$^c$ 		
		Centro de Tecnolog\'{i}as F\'{i}sicas: Acustica, Materiales y Astrof\'{i}sica,
		MALTA-Consolider Team,
		Universitat Politecnica de Valencia, 
		Valencia,
		Spain.
		\vspace{1mm}
\\
$^d$ 		
		Institute of Earth Sciences Jaume Almera, 
		MALTA Consolider Team, 
		CSIC, 
		08028 Barcelona, 
		Spain.
		\vspace{1mm}
\\
$^e$  	CELLS-ALBA Synchrotron Light Facility,
		08290 Cerdanyola del Vall\`{e}s,
		Barcelona, 
		Spain.
		\vspace{1mm}
\\
$^f$ 		
		Department of Rare Earths, 
		Faculty of Chemistry, 
		Adam Mickiewicz University, 
		Grunwaldzka 6, 
		60-780 Pozna\'{n}, 
		Poland.
		\vspace{1mm}

$^{*}$e-mail: robin.turnbull@uv.es

\end{center}

\newpage
\clearpage

\noindent{\bf Abstract}
\vspace{\baselineskip}

\begin{abstract}

\noindent
	
	High-pressure X-ray diffraction studies on nanocrystals of the pseudo-vaterite-type borates GdBO$_{3}$ and YBO$_{3}$ are herein reported up to 17.4(2) and 13.4(2) GPa respectively.
	The subsequent determination of the room-temperature pressure-volume equations of state is presented and discussed in the context of contemporary publications which contradict the findings of this work.
	In particular, the isothermal bulk moduli of GdBO$_{3}$ and YBO$_{3}$ are found to be  $\bm{170(13)}$ and $\bm{163(13)}$ GPa respectively, almost 50\% smaller than recent findings.
	Our experimental results provide an accurate revision of the high-pressure compressibility behaviour of GdBO$_{3}$ and YBO$_{3}$ which is consistent with the known systematics in isomorphic borates and previous \textit{ab initio} calculations.
	Finally we discuss how experimental/analytical errors could have led to unreliable conclusions reported elsewhere.
		
\end{abstract}

\newpage
\clearpage







\section{Introduction}

	Pseudo-vaterite rare-earth orthoborates exhibit useful luminescent properties when doped with lanthanide ions.
	They demonstrate a strong chemical stability and optical damage resistance making them desirable for optics applications\cite{Das2016,Zhang2012,Choi2011,Dong2010,Jia2010,Wei2002}.
	The luminescent properties, such as intensity and chromaticity, can be tuned via the external control parameters of pressure\cite{Song2015}.
	A natural extension of such high-pressure studies is to investigate the isothermal pressure-volume compressibility, or its inverse, the fundamental thermodynamic parameter of the isothermal bulk modulus.

	The primary objective of this work is to clarify the bulk modulus of GdBO$_{3}$ 
	by performing high pressure powder X-ray diffraction on nanocrystal samples.
	YBO$_{3}$ nanocrystals were also investigated for comparison.
	Certainly, single crystal diffraction measurements would have facilitated the refinement of atomic positions, however the structure of pseudo-vaterite orthoborates like GdBO$_{3}$ and YBO$_{3}$ is already well known\cite{Lin2004,Szczeszak2012}. 
	For the purpose of determining lattice parameters as a function of pressure, and subsequently the isothermal bulk modulus, powder XRD measurements at high pressure are more than suitable.
	The choice of nano-crystalline samples offers a route to optimise the polycrystallinity of the sample (see Supplementary Fig.  \ref{supfigXRDraw_nano_micro}) without affecting the observed compressional behaviour, as discussed below.
	N.B.: When no `nano-' or `micro-' prefix is stated, the article refers to the nanocrystal samples which are the focus of this work.

	The bulk modulus of a material is a fundamental thermodynamic parameter.
	The secondary objective of this paper is, more generally, to draw attention to the fact that the accurate and reliable determination of compressibility values requires careful data acquisition and analysis which is not always performed.
	Recently, the bulk modulus of GdBO$_{3}$ was reported by Wo\'{z}ny \textit{et al.} in Ref. \citeonline{Wozny2020} to be 326 GPa, 
	which the authors allege supports the earlier reported value of 321 GPa by Wang \textit{et al.} in Ref. \citeonline{Wang2014}.
	In our previous work, Ref. \citeonline{Turnbull2020}, we commented on the reliability of the bulk modulus alleged in Ref. \citeonline{Wang2014} pertaining to the isomorphic borate YBO$_{3}$, and the reader is referred there for more details.
	The focus of the current article is the reported bulk moduli of GdBO$_{3}$, although nanocrystal YBO$_{3}$ samples were investigated in parallel for comparison.
	The experimental results of Ref. \citeonline{Wang2014} pertaining to GdBO$_{3}$ attracted close criticism from  Ref. \citeonline{Errandonea2014}, which provided \textit{ab initio} calculations and arguments based on the systematics of isomorphic borates to suggest that the actual isothermal bulk modulus of GdBO$_{3}$ is in fact closer to 135 GPa.
	In this work, through careful data acquisition and analysis, whereby non-hydrostatic data are identified and disregarded, the bulk modulus of GdBO$_{3}$ was determined to be 170(13) GPa, which is in good agreement expected values based on compressibility systematics of orthoborates and with the predicted value of 135 GPa of Ref. \citeonline{Errandonea2014}, and is approximately 50\% of the overestimated value of 326 GPa of Ref. \citeonline{Wozny2020}, and of 321 of Ref. \citeonline{Wang2014}.

\comment{The main focus of Wo\'{z}ny \textit{et al.} in Ref. \citeonline{Wozny2020} is the high-pressure luminescence properties of GdBO$_{3}$:Eu$^{3+}$, however, the first section of the paper discusses the high-pressure structural properties of GdBO$_{3}$:Eu$^{3+}$, in particular the bulk modulus, which they allege to be 326 GPa, and which they state contradicts the \textit{ab initio} calculations of Errandonea \textit{et al.}\cite{Errandonea2014} and supports the findings of Wang \textit{et al.}\cite{Wang2014}.
	(Similarly, and as discussed in our previous work, Ref. \citeonline{Turnbull2020}, the authors of Ref. \citeonline{Wang2014} allege a YBO$_{3}$ bulk modulus of 321 GPa.)
	The pressure-volume data of Refs. \citeonline{Wozny2020} and \citeonline{Wang2014} are displayed in comparison to our own data in Fig. \ref{fig2}.}

\newpage
\clearpage

\newpage

\section{Methods}

\subsection{\textit{2.1 Sample preparation}}~\
	
	Undoped nanocrystals of monoclinic YBO$_{3}$ and GdBO$_{3}$ were sythnesised via a sol-gel Pechini method using 900 $^{\circ}$C annealing as reported in Ref. \citeonline{Szczeszak2012}.

\subsection{\textit{2.2 Measurements}}~\
	
	GdBO$\bm{_{3}}$ and YBO$\bm{_{3}}$ nanocrystals were loaded into membrane-driven diamond anvil cells (DACs) to achieve gigapascal (GPa) pressures.
	Pure Cu powder was included in an isolated sample area for use as a pressure gauge in the X-ray diffraction (XRD) experiments\cite{Dewaele2004}.
	Tungsten gaskets were pre-indented to 30 $\mu$m prior to loading the nanocrystals, and diamond anvils with culet sizes $\sim$ 300 $\mu$m were used.
	The pressure transmitting medium (PTM) was a 16:3:1 methanol-ethanol-water mixture\cite{Klotz2009}.

	Angle-dispersive synchrotron powder XRD data were acquired at ALBA Synchrotron\cite{Fauth2013} (Barcelona, Spain) on the BL04 - MSPD beamline at using a monochromatic beam $\lambda$ = 0.4246 \AA\ focused to a spot size of 20 $\times$ 20 $\mu$m. 
	A SX165 Rayonix Mar CCD detector was used to record the data.
	The nanocrystal samples were typically rotated about the axis perpendicular to the X-ray beam over a range of $\pm3^\circ$
	with a typical acquisition time of 10 s.
	The XRD patterns were masked and integrated in Dioptas\cite{Prescher2015}.
	Refinement of the calculated Le Bail profiles against the observed data was performed in JANA2006\cite{JANA2006}.
	The lattice parameteres of the refined Le Bail were then used to calculate the unit cell volumes of YBO$_{3}$ and GdBO$_{3}$ as functions of increasing pressure.
	Equations of state (EoS) were fitted to the volume-pressure data using using EoSFit7\cite{Gonzalez2016} using second-order ($B_{0}^{'} = 4$) Birch-Murnaghan equations\cite{Birch1947}, the validity of which was checked via the gradient of associted $F_{E}$ \textit{vs}. $f_{E}$ plots\cite{Angel2000} provided in the supplementary material.

\newpage
\clearpage	
	
	\section{Results}
	\subsection{\textit{3.1 X-ray crystallography and bulk moduli}}~\

\begin{figure}
\centering
\includegraphics[width=1\textwidth]{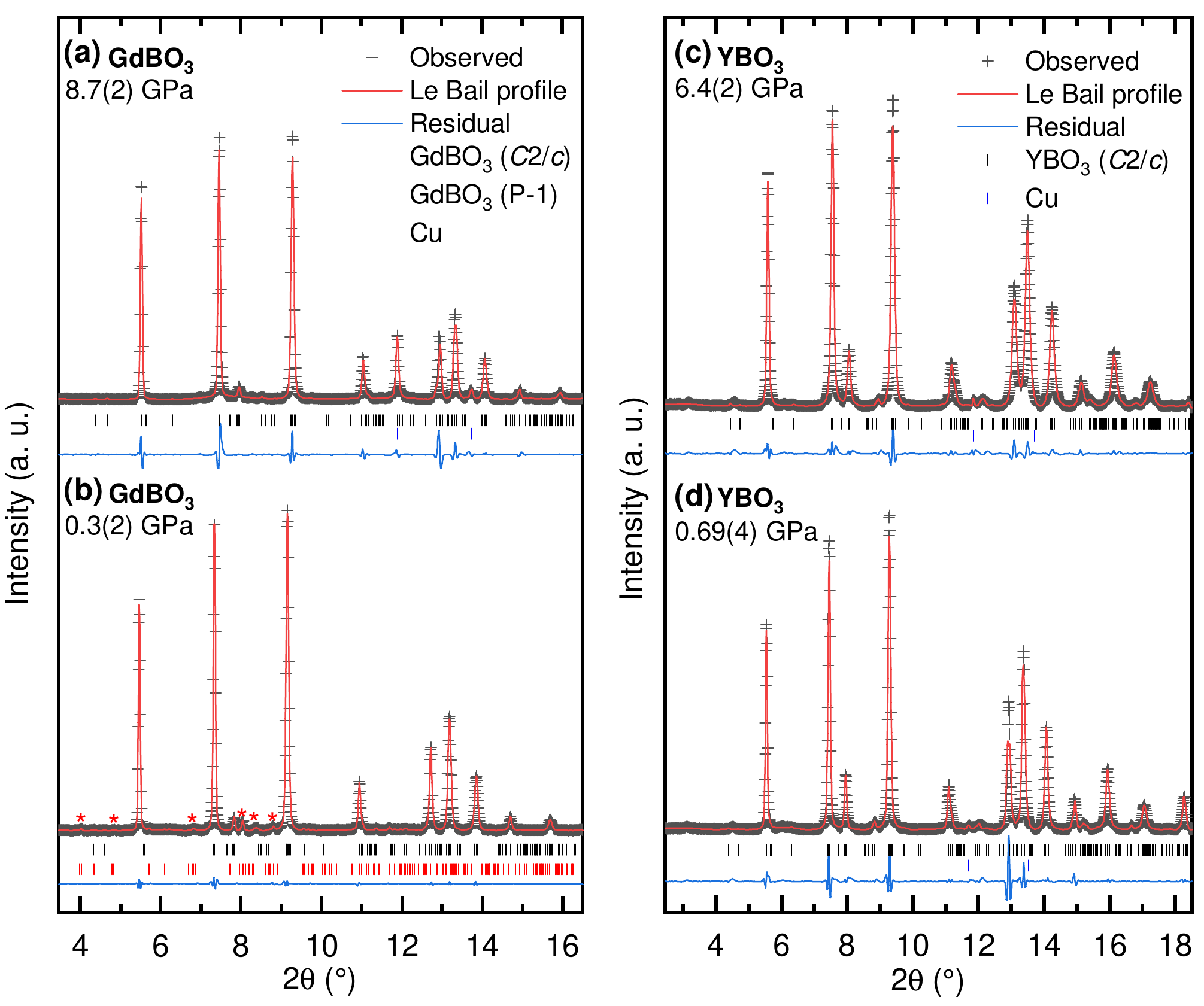}
\linespread{1.6}\caption{\textbf{Integrated XRD patterns for (a-b) nano-GdBO$_{3}$ and (c-d) nano-YBO$_{3}$.}
	Observed data points are shown with black crosses.
	The calculated Le Bail profiles are shown with red lines.
	The difference between the observed and calculated profiles is shown in blue.
	Tick marks below the profiles correspond to reflections from the compounds indicated in the legends.
	In the nano-GdBO$_{3}$ pattern at lowest pressure, \textbf{(b)}, the extremely low-intensity reflections from the triclinic phase are labelled with red asterisks.
	The lattice parameter data for the GdBO$_{3}$ and YBO$_{3}$ over the full pressure range are provided in Supplementary Tables \ref{table:nano_YBO3} and \ref{table:nano_GdBO3}
	and they are plotted individually in Supplementary Fig. \ref{supfig2}.
	The lattice parameters for the triclinic ($P-1$) GdBO$_{3}$ only observed at the lowest pressure are: $a = 6.4820(9)$, $b = 6.4682(7)$, $c = 6.2589(5)$ \AA\ and $\alpha = 108.37(1)$, $\beta = 107.00(1)$ and  $\gamma = 93.15(1) ^{\circ}$.
	}\label{fig:gdbo3XRD}
\end{figure}

	The GdBO$\bm{_{3}}$ and YBO$\bm{_{3}}$ nanocrystals were compressed at ambient temperature up to 17.4(2) and 13.4(2) GPa respectively. 
	(The numbers in parentheses are the standard errors in the least significant digit.)
	Representative integrated XRD patterns for both compounds are provided in Fig. \ref{fig:gdbo3XRD}.
	All XRD patterns are consistent with the monoclinic pseudo-vaterite $C2/c$ structure previously determined via neutron diffraction\cite{Lin2004} for YBO$_{3}$, and subsequently confirmed for GdBO$_{3}$ via X-ray diffraction\cite{Szczeszak2012}.
	No phase transition is observed in either compound over the full pressure range.
	The lattice parameters and residual values for both compounds at each pressure increment are provided in Supplementary Tables \ref{table:nano_YBO3} and \ref{table:nano_GdBO3}.
	Example raw diffraction images of nano-GdBO$_{3}$ and nano-YBO$_{3}$ are shown in Supplementary Fig. \ref{supfigXRDraw}.
	We note that although the $C2/c$ space group for the pseudo-vaterite borates was unambiguously determined via neutron diffraction\cite{Lin2004}, numerous subsequent articles have continued to use older, although similar, structural models which can now be discarded, in particular the $P6_{3}/m$, $P6_{3}/mmc$ or $P6_{3}/mcm$ space groups which do not account for a number of low intensity low-angle reflections observed via synchrotron XRD.


	The lowest pressure GdBO$\bm{_{3}}$ XRD pattern (Fig. \ref{fig:gdbo3XRD}b) indicated trace amounts of a triclinic GdBO$_{3}$ phase which were detected via extremely low-intensity low-angle reflections (marked with red asterisks).
	The triclinic GdBO$_{3}$ phase is well documented in Ref. \citeonline{Szczeszak2012}. 
	The trace impurity was not detectable using in-house XRD techniques and it is present in such small amount that it does not affect subsequent compressional data analysis of the pure monoclinic phase.
	Indeed the reflections from the triclinic GdBO$_{3}$ phase are not observable at higher pressures even with the high-sensitivity of synchrotron XRD.

\begin{figure}
\centering
\includegraphics[width=1\textwidth]{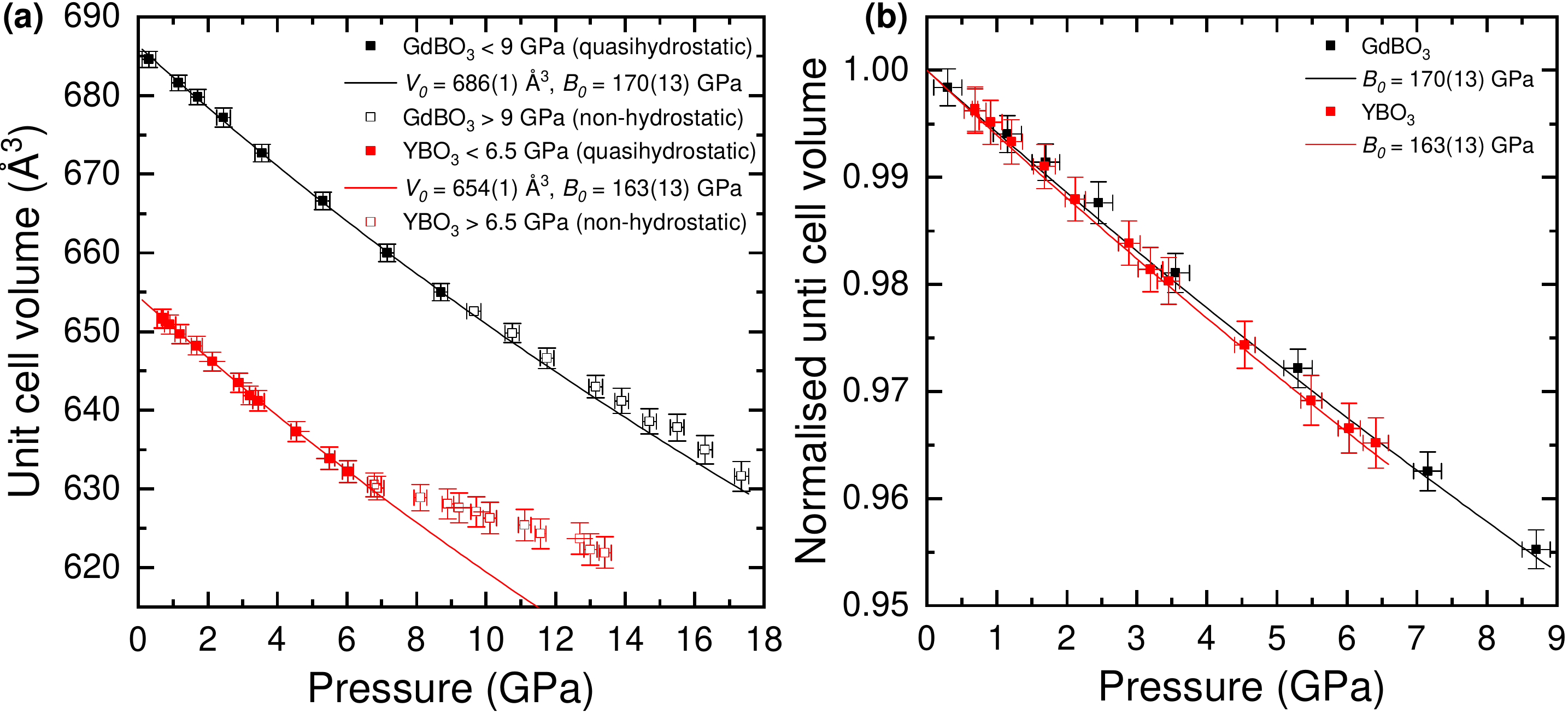}
\caption{\textbf{GdBO$_{3}$ and YBO$_{3}$ unit cell volume as function of increasing pressure.}
	Black symbols correspond to GdBO$_{3}$ and red symbols to YBO$_{3}$.
	Solid symbols correspond data acquired under quasihydrostaic conditions, and empty symbols to data acquired under non-hydrostaic conditions.
	The lines correspond to second-order Birch-Murnaghan equations of state.
	The error bars were determined from the Le Bail refinements of the powder X-ray diffraction patterns.
	The pressure was determined from the copper equation of state.
	}\label{fig1}
\end{figure}

	The integrated XRD patterns for GdBO$\bm{_{3}}$ and YBO$\bm{_{3}}$ shown in Fig. \ref{fig:gdbo3XRD} are of the lowest and highest pressures observed within the quasihydrostatic pressure range.
	The onset of non-hydrostaticity in the GdBO$_{3}$ and YBO$_{3}$ samples was identified by observing the evolution of the full-width half maximum (FWHM) of the ($002$) reflection with increasing pressure (shown in Supplementary Fig. \ref{supfig1}).
	The clear discontinuity in the rate of change of FWHM at 6.5 GPa for YBO$_{3}$ and 9 GPa for GdBO$_{3}$ indicate the loss of quasihydrostaticity and all data above these pressures have been omitted from further analysis, although they are still included in Fig. \ref{fig1}a to illustrate that the onset of non-hydrostaticity is also observed in the calculated unit cell volumes as a function of pressure.
	Integrated XRD patterns for GdBO$_{3}$ and YBO$_{3}$ at the maximum investigated pressures in non-hydrostatic range, 17.4(2) and 13.4(2) GPa respectively, are shown in Supplementary Fig. \ref{XRD_waterfall_maxP}.

	The unit cell volumes of GdBO$_{3}$ and YBO$_{3}$, determined by Le Bail analysis, are plotted in Fig. \ref{fig1} as a function of increasing pressure.
	As shown in Fig. \ref{fig1}a, second-order Birch-Murnaghan (BM) equations of state (EoS) provide  excellent fits to the data below the aforementioned quasihydrostatic limits, providing isothermal bulk moduli\cite{Angel2000} of $B_{0} =170(13)$ and $163(13)$ GPa for GdBO$_{3}$ and YBO$_{3}$ respectively,
	where the bulk modulus, $B$, is given by, $B = -V \delta P / \delta V$.
	The data above the quasihydrostatic limit (shown with empty symbols) clearly diverge from the second-order BM EoS.
	Using the same input data and type of EoS, the PASCal principal axis strain calculator\cite{Cliffe2012} calculates very similar bulk moduli of 170(2) and 163(1) GPa for dBO$_{3}$ and YBO$_{3}$ respectively.
	Fig. \ref{fig1}b shows the normalised unit cell volume for both compounds, emphasising the quasihydrostatic data only.
	The EOS fits were constrained to second-order ($B_{0}^{'} = 4$) to limit the number of fitting parameters and to thereby facilitate comparison.
	The $F_{E}$ \textit{vs}. $f_{E}$ plots provided Supplementary Fig. \ref{fig:ff-plot} provide a useful visual assessment of the quality of the fitted equations of state.
	The excellent suitability of the second-order EOS truncation is demonstrated via the essentially zero gradient in those plots.
	The reader is referred to Ref. \citeonline{Angel2000} for more details regarding 
	$F_{E}$ \textit{vs}. $f_{E}$ plots.
	In addition to the bulk modulus of GdBO$_{3}$, we also calculated for the first time the individual axial compressibilities via the isothermal compressibility tensor\cite{Knight2010} (see Supplementary Table \ref{table:tensor}) which reveals an anisotropic compressibility in GdBO$_{3}$ essentially identical to that exhibited by YBO$_{3}$ in our previous work\cite{Turnbull2020} and which we rationalised in terms of the compressibilities of the constituent BO$_{4}$-tetrahedra and $A$O$_{8}$-dodecahedra.

\newpage
\clearpage

\newpage
\clearpage

\subsection{\textit{3.2 Critical comparison of bulk moduli with reported values}}
\begin{figure}
\centering
\includegraphics[width=1\textwidth]{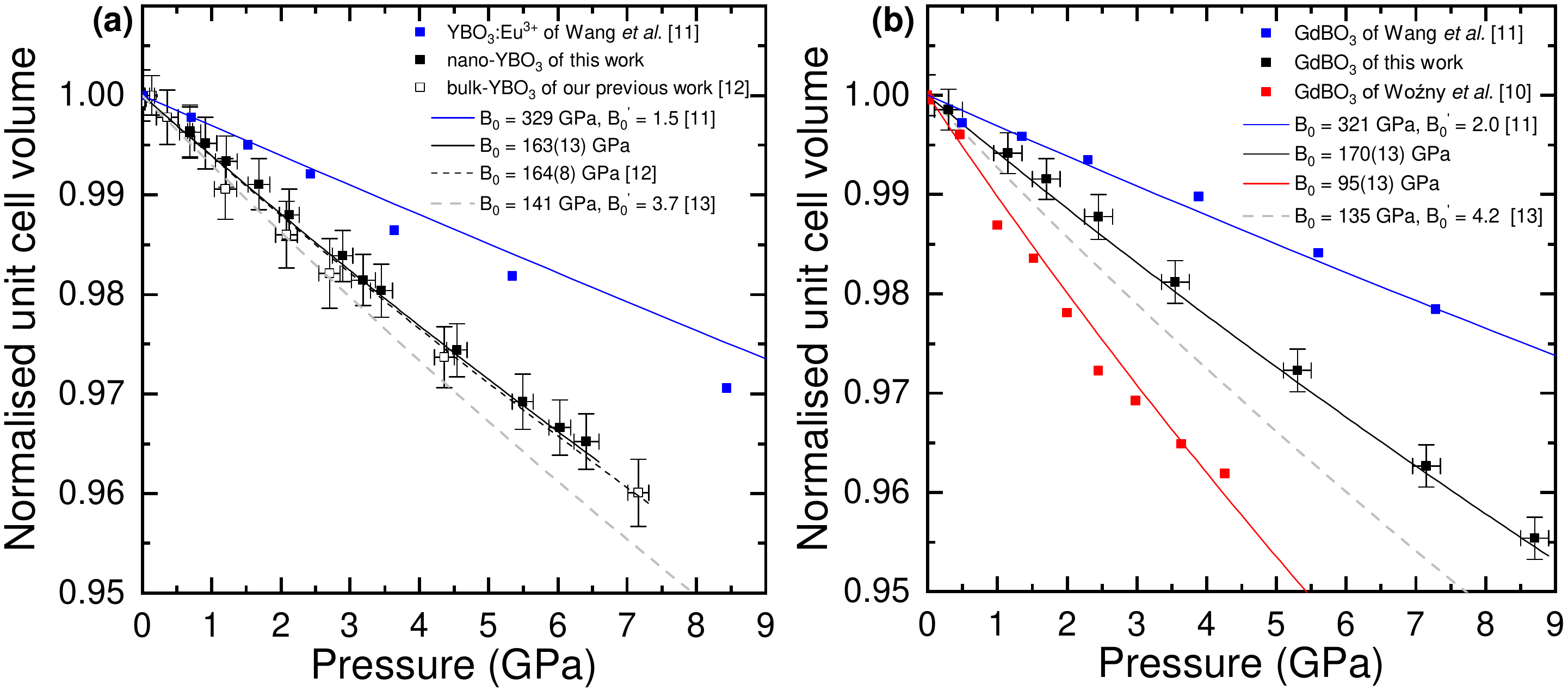}
\caption{\textbf{Comparison of the reported normalised volumes of (a) YBO$_{3}$ and (b) GdBO$_{3}$.}
	The black data correspond to data acquired by the authors of this article here and in Ref. \citeonline{Turnbull2020}.
	Blue data are reported by Wang \textit{et al.} in Ref. \citeonline{Wang2014}.
	Red data are reported by Wo\'{z}ny \textit{et al.} in Ref. \citeonline{Wozny2020}.	
	Lines correspond to second-order Birch-Murnaghan equations of state when $B_{0}^{'}$ is not stated.
	The blue EoS are reported in Ref \citeonline{Wang2014}.
	The red EoS was fitted by the present authors with a drastically lower bulk modulus (95(13) GPa) than reported in Ref. \citeonline{Wozny2020} (326.09(4) GPa).
	The dashed grey EoS are from the \textit{ab initio} calculations of Ref. \citeonline{Errandonea2014}.
 	}\label{fig2}
\end{figure}

~\\
	Two important points must be made before commencing with the following discussion.
	Firstly, we are able to make a direct comparison between the compressional properties of GdBO$_{3}$ of this work, and the GdBO$_{3}$:Eu$^{3+}$ of Ref. \citeonline{Wozny2020} because eightfold-coordinated Gd$^{3+}$ and Eu$^{3+}$ ions have very similar ionic radii of 1.053 and 1.066 \r{A}, respectively, according Ref. \citeonline{Shannon1976}. 
	Therefore, the bulk moduli of GdBO$_{3}$ and GdBO$_{3}$:Eu$^{3+}$ can be expected to be essentially identical.
	Secondly, we are also able to make a direct comparison between powder samples consisting of nano-crystals and micro-crystals.
	This is shown in Fig. \ref{fig2}a where the nano-YBO$_{3}$ data of this work are compared with the micro-YBO$_{3}$ data of our previous work, Ref. \citeonline{Turnbull2020}.
	The bulk modulus of the micro-YBO$_{3}$ samples in our previous work was determined to be 164(8) GPa.
	This is in excellent agreement with the bulk modulus determined here for the nano-YBO$_{3}$ sample of 163(13) GPa.
	The nano-YBO$_{3}$ and micro-YBO$_{3}$ data in Fig. \ref{fig2}a perfectly agree to within the displayed errors.
	(The data of Wang \textit{et al.} Ref. \citeonline{Wang2014} are included in Fig. \ref{fig2}a for comparison, a detailed discussion of which is available in Ref. \citeonline{Turnbull2020}.)

	The main focus of Wo\'{z}ny \textit{et al.} in Ref. \citeonline{Wozny2020} is the high-pressure luminescence properties of GdBO$_{3}$:Eu$^{3+}$,
	however,
	section 3.2 of the paper discusses the high-pressure structural properties of GdBO$_{3}$:Eu$^{3+}$, in particular the bulk modulus, which they allege to be 326 GPa, and which they state contradicts the \textit{ab initio} calculations of Errandonea \textit{et al.}\cite{Errandonea2014} and supports the findings of Wang \textit{et al.}\cite{Wang2014}.
	We limit our critique of Ref. \citeonline{Wozny2020} to the results presented in their section 3.2.
	The pressure-volume data of Refs. \citeonline{Wozny2020,Wang2014,Turnbull2020,Errandonea2014}
	are displayed in comparison the data of this work in Fig. \ref{fig2}.
	
\newpage
\clearpage

	In Fig. \ref{fig2}b the GdBO$_{3}$ data of the present work, shown in black, are very well fitted with a bulk modulus of 170(13) GPa.
	The data of Wang \textit{et al.}, shown in blue, are well described by their alleged equations of state, however for reasons previously discussed (see Ref. \citeonline{Turnbull2020}) their bulk moduli for YBO$_{3}$ and GdBO$_{3}$ are likely to be an overestimation by a factor of approximately 2.
	The data of Wo\'{z}ny \textit{et al.}, shown in red, have here been refitted with a bulk modulus of 95(13) GPa.
	There are three important remarks:
	firstly that the alleged GdBO$_{3}$ bulk modulus of 326 GPa is not compatible with their data; 
	secondly, that in any case such a large bulk modulus is not compatible with the known compressibility systematics of isomorphic borates\cite{Santamaria2014};
	and thirdly, that the low bulk modulus of 95(13) GPa re-fitted here drastically underestimates the bulk modulus of GdBO$_{3}$.
	Therefore, 
	the data of Wo\'{z}ny \textit{et al.} do not agree with those of Ref. \citeonline{Wang2014} (shown in blue) as they claim.
	Their result is potentially the product of problems either with their experimental data or the analysis thereof, however it is not possible to say based on the data they present, for example:
	not a single a XRD refinement is presented and only integrated patterns are provided; 	no details are provided of how the XRD fits were made;
	no details are provided of how the equations of state were fitted, such as the type or order;
	the XRD patterns exhibit gasket reflections from the very lowest pressure; and,
	the XRD reflections are broad/exhibit low-intensity at low pressures (compare for example to the data in Fig. \ref{fig:gdbo3XRD}).
	For their XRD experiments Daphne 7575 oil was used,
	which has a hydrostatic limit at ambient temperature of $\sim4$ GPa, so it is not clear why the XRD peaks are so broad at low pressure, although it is possible that the sample chamber was too densely packed to allow room for sufficient PTM, therefore leading to bridging of the sample between diamonds.
	
	Regarding the triclinic GdBO$_{3}$ phase reported on by Wozny \textit{et al.}, 
	we cannot recreate the equation of state shown in their Fig. 2c.
	Numerous different equation of state have been fitted to their data using their reported values of $V_{0} = 230.77$ \AA\ and $B_{0} = 27.09$ GPa as fixed parameters, as shown in Supplementary Fig. \ref{fig:wozny_lowBphase}, however we were not able to reproduce the reported EoS. 
	Relaxing the fixed parameters results in converged fit with a very low bulk modulus of 19(3) GPa, which describes the data very closely but still raises concerns.
	For example, a bulk modulus of 19(3) GPa would suggest that the triclinic GdBO$_{3}$ compound is as compressible as a rare-gas solid, which cannot be the case.
	This points to unexplained experimental errors which have lead to underestimations in both monoclinic and triclinic GdBO$_{3}$ bulk moduli. 
	The essential difference between the triclinic and monoclinic GdBO$_{3}$ phases is that the BO$_{4}$-tetrahedra in the monoclinic phase open to up form tringular BO$_{3}$-units in the triclinic phase. 
	Therefore, although a lower bulk modulus is indeed expected for the triclinic phase,  the fact that the triclinic structure contains GdO$_{8}$-octahedra and BO$_{3}$-trigonal-units certainly indicates a compound more rigid than a rare gas solid.	
	Therefore, further high-pressure experiments are required on the triclinic GdBO$_{3}$ structure in order to clarify its isothermal compressional behaviour.


\newpage
\clearpage	
	
	\section{Conclusions}

	This experimental X-ray diffraction study of GdBO$_{3}$ and YBO$_{3}$ nanocrystals under compression, up to 17.4(2) and 13.4(2) GPa respectively, has provided an accurate revision of the compressional behaviour of monoclinic GdBO$_{3}$.
	In particular, the fundamental property of bulk modulus	was found to be 170(13) and 163(13) GPa for GdBO$_{3}$ and YBO$_{3}$ nanocrystals respectively, thereby revealing that the compressional behaviour of both compounds is consistent with the family of borate compounds.
	The \textit{ab initio} calculations of Ref. \citeonline{Errandonea2014} predicted a GdBO$_{3}$ bulk modulus of 135 GPa, which consistent with the results presented here, and those expected based on observed compression systematics based on metallic cationic radius\cite{Santamaria2014}.
	The previously alleged bulk moduli\cite{Wozny2020,Wang2014} suggested that GdBO$_3$ would be less compressible than all known ultra-incompressible nitride and carbides\cite{Errandonea2010} which is not expected for a layered structure like that of vaterite\cite{Mugnaioli2012}.		
	This study concludes that the claims of GdBO$_{3}$ incompressibility made by other authors are not correct, in particular that the GdBO$_{3}$ bulk modulus values of 326 GPa reported in Ref. \citeonline{Wozny2020}, and of 321 GPa in Ref. {\citeonline{Wang2014}}, are overestimated by a factor of approximately 2.
	In this work, comparison of the compressional behaviour of GdBO$_{3}$ with structurally analogous YBO$_{3}$ provides very similar results for both  compounds as expected.
	The YBO$_{3}$ bulk modulus determined here (163(13) GPa) agrees very closely with our previous result\cite{Turnbull2020} on YBO$_{3}$ micro-powders of 164(8) GPa, and shows that in the case of pseudo-vaterite borates optimisation of powder XRD experiments is possible through the use of nano-powder samples without affecting the observed compressional behaviours.
	The results of this work highlight that careful data acquisition and analysis are necessary for the accurate and reliable determination of the fundamental thermodynamic value of the isothermal bulk modulus.

\newpage
\clearpage

\noindent\textbf{Data availability} ---
	All relevant data are available from the corresponding author upon reasonable request.
\\

\noindent\textbf{Acknowledgements} ---
	The authors thank the financial support from the Spanish \textit{Ministerio de Ciencia, Innovacion y Universidades}, Spanish  Research Agency (AEI), \textit{Generalitat Valenciana}, and European Fund for Regional Development (ERDF, FEDER) under grants no. 
	F152017-83295-P, 
	MAT2016-75586-C4-1/2/3-P, 
	RTI2018-101020-BI00,
	PID2019-106383  GB-C41/C42/C43, 
	RED2018-102612-T (MALTA Consolier Team), 
	and Prometeo/2018/123 (EFIMAT). 
	R.T. acknowledges funding from the Spanish MINECO via the \textit{Juan de la Cierva Formación} program (FJC2018-036185-I),
	and J.Á.S. acknowledges funding from the \textit{Ramón y Cajal} fellowship program (RYC-2015-17482). 
	We also thank ALBA synchrotron light source for funded experiments 2016021648 and 2016021668 at the MSPD-BL04 beamline.\\

\noindent{\bf References}


\comment{

\noindent\textbf{Online Content} Supplementary display items are available in the online version of the paper.

\noindent\textbf{Acknowledgements}
	The authors thank the financial support from the Spanish Ministerio de Ciencia, Innovacion y Universidades, Spanish  Research Agency (AEI), Generalitat Valenciana, and European Fund for Regional Development (ERDF, FEDER) under grants no. MAT2016-75586-C4-1/2/3-P, RTI2018-101020-BI00, RED2018-102612-T (MALTA Consolier Team), and Prometeo/2018/123 (EFIMAT). 
	R.T. acknowledges funding from the Spanish MINECO via the Juan de la Cierva Formación program (FJC2018-036185-I).

\noindent 
	We acknowledge ALBA/DLS for provision of synchrotron radiation facilities at the X beamline. 
	Parts of this research were carried out under proposal XXX.

\noindent\textbf{Author Contributions}
	D.E. and F.J.-M. conceived and designed the project. 
	F.J.-M., V.P.C.-G., J.Á.S., R.V. and C.P. conducted the experiments.
	R.T., D.E. and K.K.M. analysed the data.
	R.T., D.E., A.M.-G., P.R.-H. and M.B. wrote the paper.
	A.M.-G. and P.R.-H. performed the calculations.
	M.B. Synthesised the YBO$_{3}$ crystals.

\noindent\textbf{Author Information}
	Readers are welcome to comment on the online version of the paper. 			
	Correspondence and requests for materials should be addressed to R.T. (robin.turnbull@uv.es).

\noindent\textbf{Competing Interests}
	The authors declare no competing interests.

}

\newpage
\clearpage

\setcounter{figure}{0}
\captionsetup[figure]{labelfont={bf},name={Supplementary Figure},labelsep=pipe}
\setcounter{table}{0}
\captionsetup[table]{labelfont={bf},name={Supplementary Table},labelsep=pipe}

\LARGE{Supplementary Material}

\newpage
\clearpage

\begin{figure}
\centering
\includegraphics[width=1\textwidth]{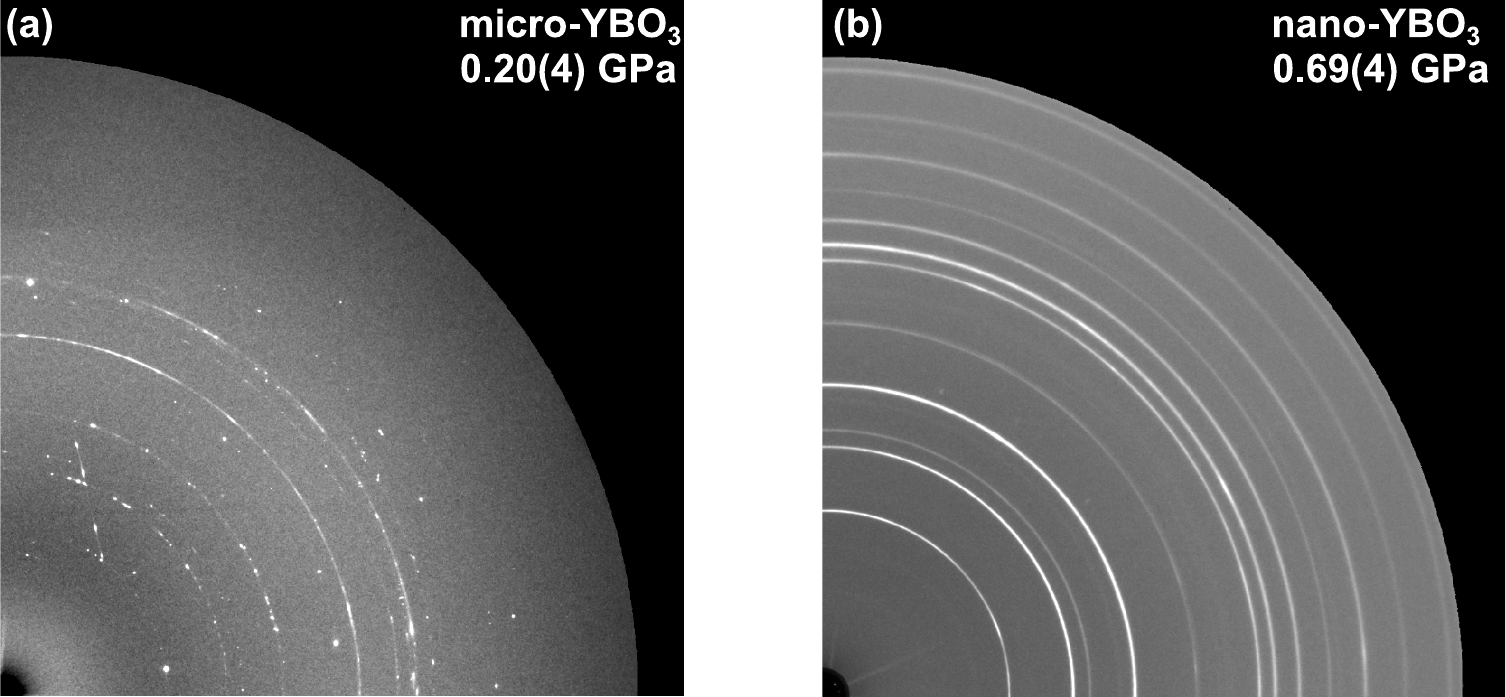}
\caption{\textbf{Raw diffraction patterns.}
	\textbf{(a)} micro-YBO$_{3}$ at 0.69(4) GPa, as used in Ref. \citeonline{Turnbull2020}.
	\textbf{(b)} nano-YBO$_{3}$ at 0.69(4) GPa, used in this work.
	}\label{supfigXRDraw_nano_micro}
\end{figure}

\newpage
\clearpage

\renewcommand{\arraystretch}{0.7}
\begin{table}[t]
\small
\centering
\begin{tabular}{rllllll}
\hline
\hline
$P$ (GPa) & \multicolumn{1}{l}{$a$ (\r{A})} & \multicolumn{1}{l}{$b$ (\r{A})} & \multicolumn{1}{l}{$c$ (\r{A})} & \multicolumn{1}{l}{$\beta$ (\r{})} & $V$ (\r{A}$^{3}$) & $R_{p}$\\
0.69(4) & 11.2885(9) & 6.5671(5) & 9.5427(7) & 112.900(6) & 651.7(12) & 24.82 \\
0.7(2) & 11.2887(9) & 6.5665(5) & 9.5424(7) & 112.900(6) & 651.6(12) & 24.73 \\
0.9(2) & 11.2823(8) & 6.5653(6) & 9.5391(7) & 112.905(6) & 650.9(12) & 24.77 \\
1.2(2) & 11.2745(8) & 6.5604(5) & 9.5347(7) & 112.891(6) & 649.7(12) & 24.83 \\
1.7(2) & 11.263(8) & 6.5563(6) & 9.5287(7) & 112.89(6) & 648.2(12) & 25.31 \\
2.1(1) & 11.2497(9) & 6.5475(6) & 9.522(7) & 112.871(6) & 646.2(12) & 25.91 \\
2.9(2) & 11.2285(9) & 6.5391(6) & 9.5109(8) & 112.862(7) & 643.5(12) & 26.52 \\
3.2(2) & 11.2166(9) & 6.5329(6) & 9.5065(8) & 112.852(7) & 641.9(12) & 26.79 \\
3.5(2) & 11.2112(9) & 6.5295(6) & 9.5041(8) & 112.845(7) & 641.2(13) & 27.07 \\
4.5(1) & 11.1812(10) & 6.5197(6) & 9.4856(8) & 112.825(7) & 637.3(13) & 29.05 \\
5.5(2) & 11.1561(10) & 6.5079(6) & 9.4708(9) & 112.804(7) & 633.9(14) & 30.23 \\
6.0(2) & 11.1434(10) & 6.5025(6) & 9.4635(9) & 112.792(7) & 632.2(14) & 31.01 \\
6.4(2) & 11.1331(10) & 6.5018(6) & 9.4596(9) & 112.776(7) & 631.3(14) & 31.91 \\
\hline             
6.8(2) & 11.1273(11) & 6.5000(7) & 9.4547(10) & 112.774(8) & 630.5(15) & 32.86 \\
6.9(2) & 11.1218(11) & 6.4996(7) & 9.4523(10) & 112.764(8) & 630.1(15) & 32.8 \\
8.1(2) & 11.1019(12) & 6.4986(7) & 9.4497(11) & 112.705(9) & 628.9(17) & 36.39 \\
8.9(2) & 11.0781(14) & 6.5032(8) & 9.4449(12) & 112.622(9) & 628.1(19) & 38.94 \\
9.2(1) & 11.0719(14) & 6.5026(8) & 9.4427(12) & 112.611(9) & 627.6(19) & 39.56 \\
9.7(2) & 11.0651(15) & 6.5024(8) & 9.4398(12) & 112.594(10) & 627.1(19) & 40.23 \\
10.1(2) & 11.0572(15) & 6.5024(8) & 9.4344(12) & 112.578(10) & 626.3(20) & 41.3 \\
11.1(2) & 11.0491(15) & 6.5024(8) & 9.4264(12) & 112.566(10) & 625.4(20) & 42.59 \\
11.6(2) & 11.0403(15) & 6.4978(8) & 9.4234(12) & 112.55(10) & 624.3(19) & 43.19 \\
12.7(1) & 11.0339(15) & 6.4968(9) & 9.4193(12) & 112.532(10) & 623.7(20) & 44.28 \\
13.0(2) & 11.0228(15) & 6.4933(9) & 9.4123(13) & 112.517(11) & 622.3(20) & 44.78 \\
13.4(2) & 11.012(15) & 6.4949(9) & 9.4091(12) & 112.466(11) & 621.9(20) & 45.26 \\
\hline
\hline
\end{tabular}
\caption{\label{table:nano_YBO3}
 		\textbf{
 		Refined lattice parameters for nano-YBO$_{3}$.}
		Numbers in parentheses are the estimated standard error in the least significant digit. 		
 		The horizontal line mid-table indicates the onset of non-hydrostaticity according to the FWHM analysis (see Supplementary Figure \ref{supfig1}).
 		}
\end{table}

\newpage
\clearpage

\renewcommand{\arraystretch}{0.7}
\begin{table}[t]
\small
\centering
\begin{tabular}{rllllll}
\hline
\hline
$P$ (GPa) & \multicolumn{1}{l}{$a$ (\r{A})} & \multicolumn{1}{l}{$b$ (\r{A})} & \multicolumn{1}{l}{$c$ (\r{A})} & \multicolumn{1}{l}{$\beta$ (\r{})} & $V$ (\r{A}$^{3}$) & $R_{p}$\\
0.3(2) & 11.4764(6) & 6.6979(5) & 9.6776(6) & 113.035(5) & 684.6(10) & 19.36 \\
1.2(2) & 11.4551(6) & 6.6884(6) & 9.6657(6) & 113.007(5) & 681.6(10) & 22.95 \\
1.7(2) & 11.4417(7) & 6.6817(6) & 9.6590(6) & 112.993(6) & 679.8(10) & 23.83 \\
2.5(2) & 11.4238(8) & 6.6718(7) & 9.6504(7) & 112.977(7) & 677.2(12) & 27.22 \\
3.6(2) & 11.3953(7) & 6.6513(6) & 9.6394(7) & 112.957(6) & 672.7(11) & 26.11 \\
5.3(2) & 11.3547(7) & 6.6283(6) & 9.6163(7) & 112.925(7) & 666.6(11) & 25.85 \\
7.2(2) & 11.3120(7) & 6.6013(6) & 9.5928(7) & 112.886(6) & 660.0(11) & 24.71 \\
8.7(2) & 11.2813(8) & 6.5809(6) & 9.5748(7) & 112.862(7) & 655.0(11) & 25.63 \\
\hline             
9.7(2) & 11.2624(8) & 6.5717(6) & 9.5664(7) & 112.824(6) & 652.6(12) & 25.94 \\
10.8(2) & 11.2401(8) & 6.5605(6) & 9.5577(8) & 112.788(7) & 649.8(12) & 26.66 \\
11.8(2) & 11.2148(9) & 6.5484(6) & 9.5470(8) & 112.754(7) & 646.6(13) & 28.99 \\
13.2(2) & 11.1823(10) & 6.5339(6) & 9.5381(9) & 112.691(7) & 643.0(14) & 35.21 \\
13.9(2) & 11.1690(12) & 6.5267(6) & 9.5330(10) & 112.679(8) & 641.2(16) & 32.34 \\
14.7(2) & 11.1423(12) & 6.5168(6) & 9.5287(10) & 112.639(8) & 638.6(16) & 36.41 \\
15.5(2) & 11.1398(13) & 6.5125(6) & 9.5241(10) & 112.618(9) & 637.8(17) & 34.31 \\
16.3(2) & 11.1111(14) & 6.5052(6) & 9.5135(10) & 112.574(9) & 635.0(18) & 37.20 \\
17.4(2) & 11.0881(14) & 6.4910(6) & 9.5027(11) & 112.563(10) & 631.6(19) & 37.81 \\
\hline
\hline
\end{tabular}
\caption{\label{table:nano_GdBO3}
 		\textbf{
 		Refined lattice parameters for nano-GdBO$_{3}$.}
		Numbers in parentheses are the estimated standard error in the least significant digit. 		
 		The horizontal line mid-table indicates the onset of non-hydrostaticity according to the FWHM analysis (see Supplementary Figure \ref{supfig1}).
 		}
\end{table}

\newpage
\clearpage

\begin{figure}
\centering
\includegraphics[width=1\textwidth]{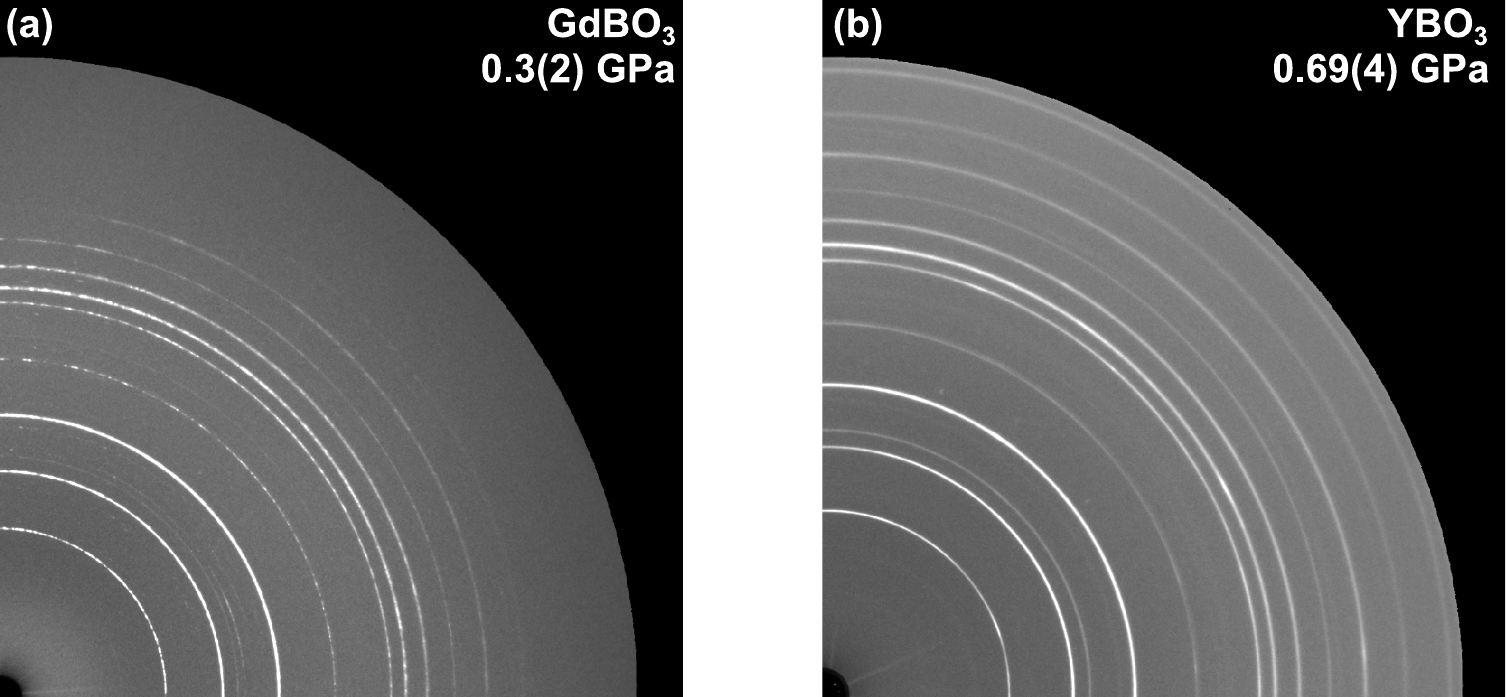}
\caption{\textbf{Raw diffraction patterns.}
	\textbf{(a)} GdBO$_{3}$ at 0.3(2) GPa.
	\textbf{(b)} YBO$_{3}$ at 0.69(4) GPa.
	}\label{supfigXRDraw}
\end{figure}

\newpage
\clearpage

\begin{figure}
\centering
\includegraphics[width=0.6\textwidth]{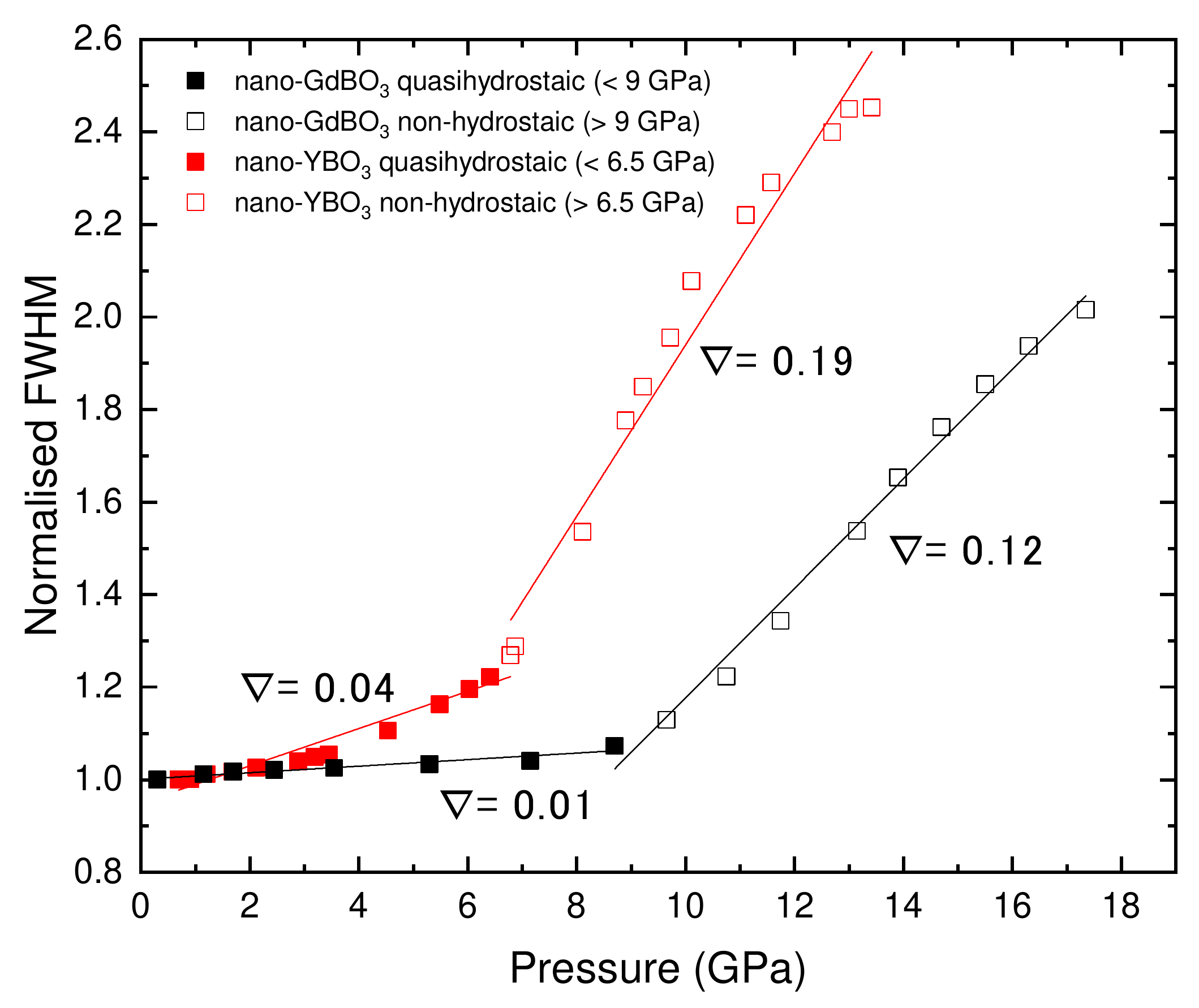}
\caption{\textbf{Normalised full-width at half maximum of the ($\bm{002}$) reflection of Gd/Y:BO$_{3}$ with increasing pressure.}
	The lines represent linear least squares fits to the quasi-hydrostatic (solid symbols) and non-hydrostatic (empty symbols) data.
 	}\label{supfig1}
\end{figure}

\newpage
\clearpage

\begin{figure}
\centering
\includegraphics[width=1\textwidth]{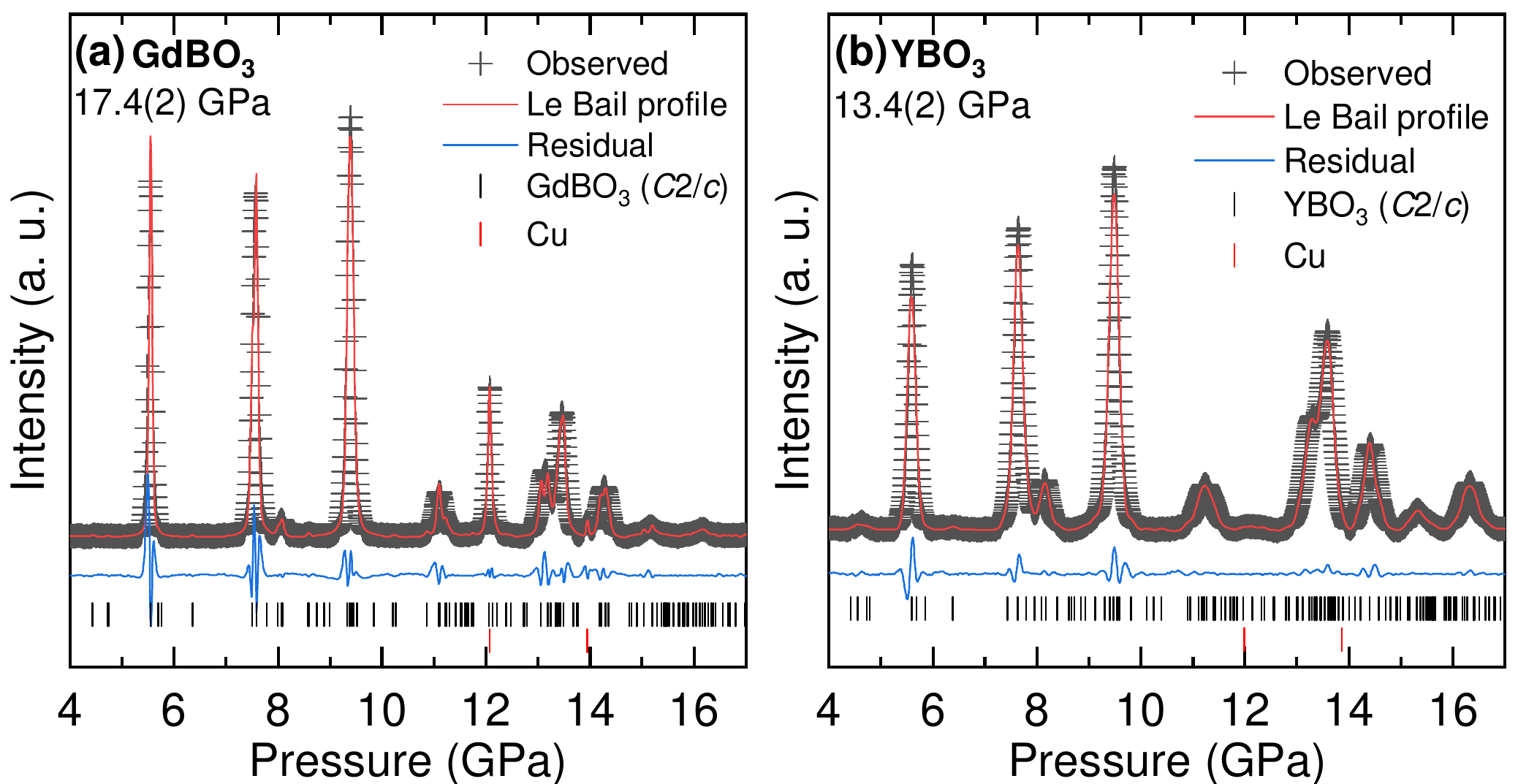}
\caption{\textbf{Integrated XRD patterns for (a) nano-GdBO$_{3}$ and (b) nano-YBO$_{3}$ at the maximum investigated pressures of 17.4(2) and 13.4(2) GPa respectively in the non-hydrostatic regime.}
	Observed data points are shown with black crosses.
	The calculated Le Bail profiles are shown with red lines.
	The difference between the observed and calculated profiles is shown in blue.
	Tick marks below the profiles correspond to reflections from the compounds indicated in the legends.
	The lattice parameter data for GdBO$_{3}$ and YBO$_{3}$ over the full pressure range are provided in Supplementary Tables \ref{table:nano_YBO3} and \ref{table:nano_GdBO3}
	and they are plotted individually in Supplementary Fig. \ref{supfig2}.
	}\label{XRD_waterfall_maxP}
\end{figure}

\newpage
\clearpage

\begin{figure}
\centering
\includegraphics[width=1\textwidth]{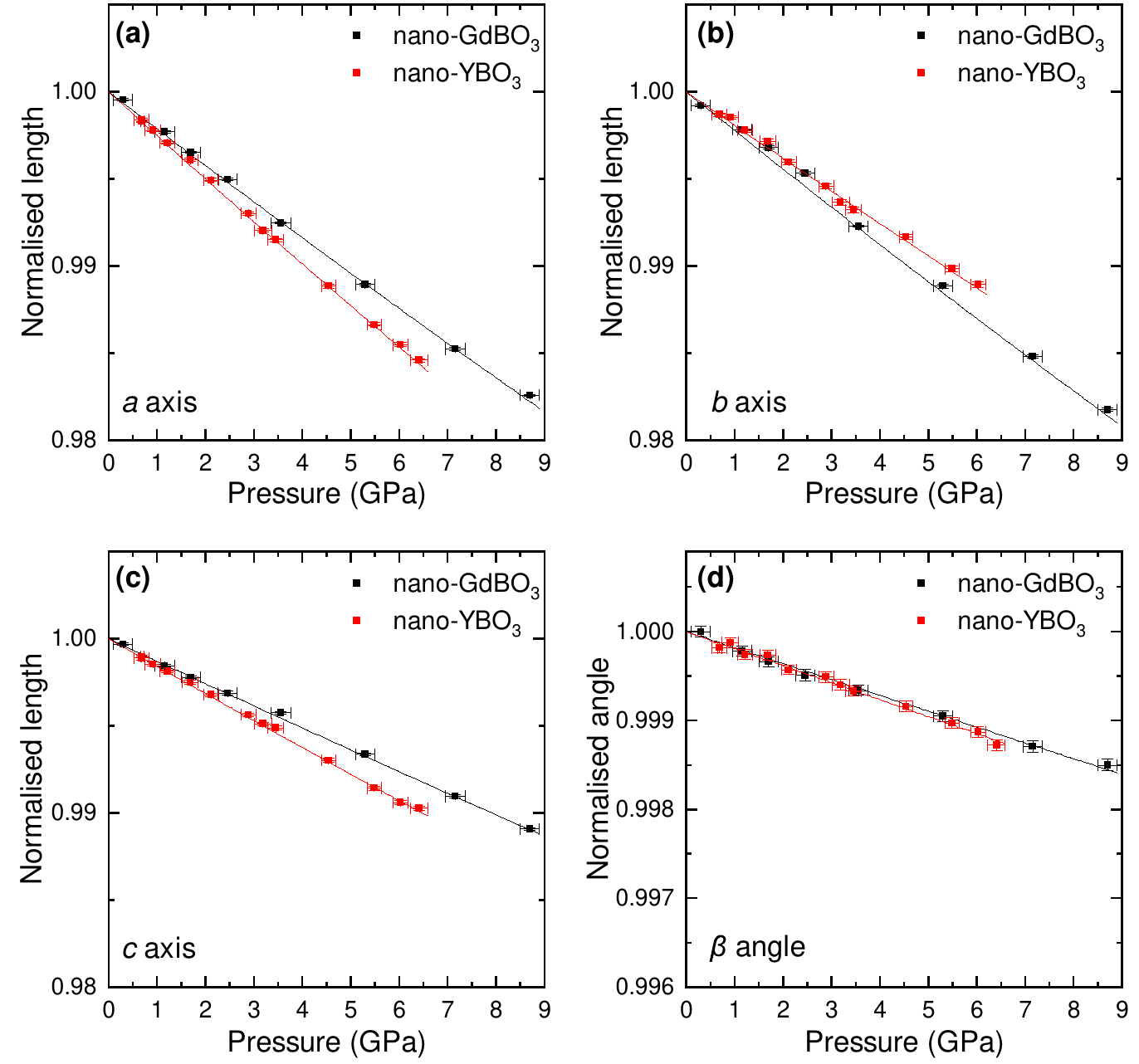}
\caption{\textbf{Normalised lattice parameters ($\bm{a}$, $\bm{b}$, $\bm{c}$ and \bm{$\beta$}) for Gd/Y:BO$_{3}$ as a function of increasing pressure.}
	The lines correspond to second-order Birch-Murnaghan equations of state.
	The lattice parameter data are provided in Supplementary Tables \ref{table:nano_YBO3} and \ref{table:nano_GdBO3}.
 	}\label{supfig2}
\end{figure}

\newpage
\clearpage

\begin{figure}[b]
\centering
\includegraphics[width=0.7\textwidth]{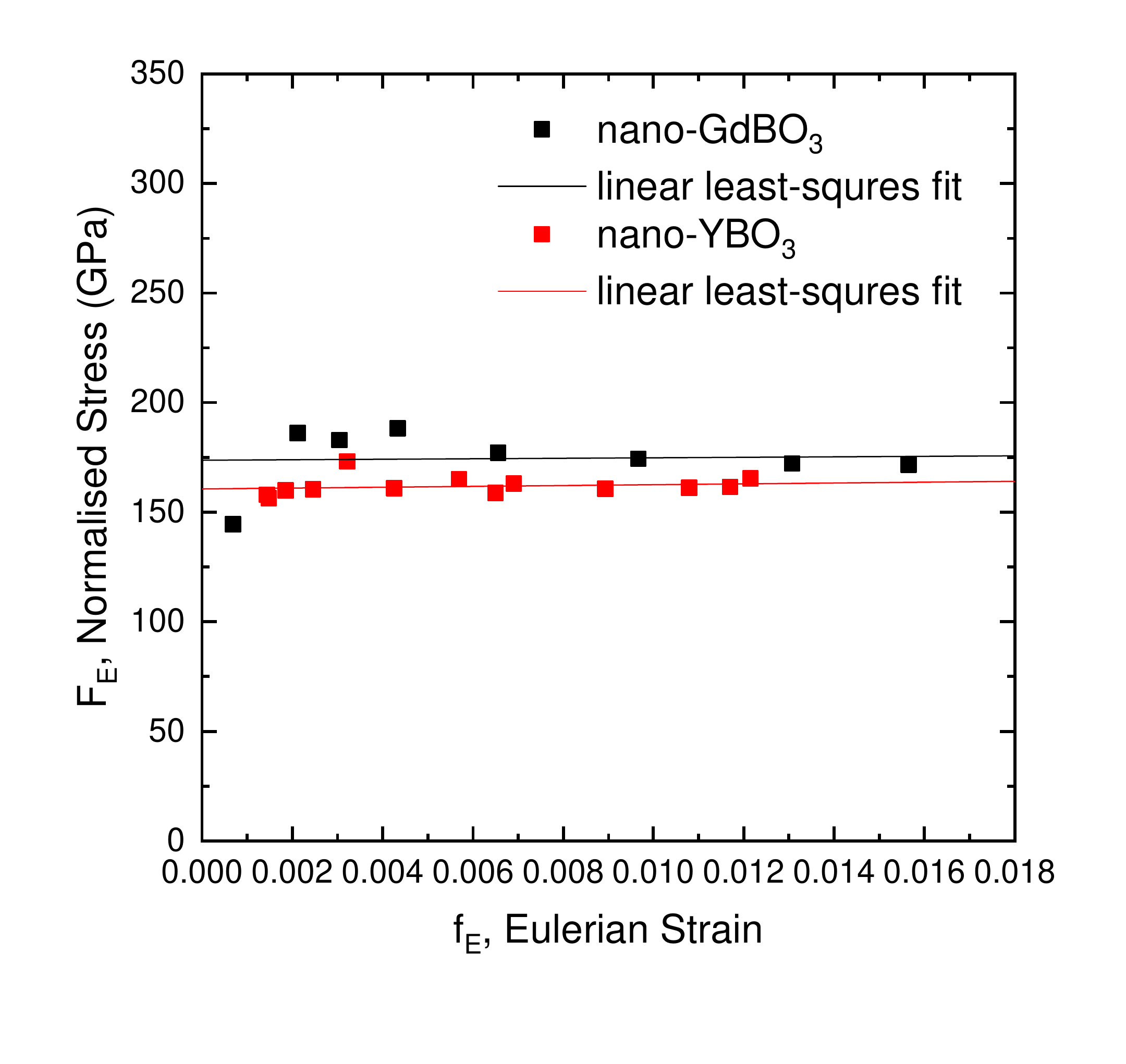}
\caption{\textbf{$\bm{F_{E}}$ \textit{vs}. $\bm{f_{E}}$ plot for the hydrostatic nano-GdBO$_{3}$ (black) and nano-YBO$_{3}$ (red) compression data.}
	The linear least squares fits to the data show essentially zero gradient, therefore
	the second order truncation of the BM EoS (i.e. $B_{0}^{'} = 4$) describes the data
	sufficiently well.
	The reader is referred to Ref. \citeonline{Angel2000} for more details regarding 
	$F_{E}$ \textit{vs}. $f_{E}$ plots.}\label{fig:ff-plot}
\end{figure}

\newpage
\clearpage

\begin{table}
\small
    \centering
     \begin{tabular}{ll}
   \hline
   \hline  
	$a_{0}$ (\r{A}), d$a/$d$P$ (\r{A} GPa$^{-1}$)  		&	$11.482(8), -0.0234(1)$\\
	$b_{0}$ (\r{A}), d$b/$d$P$ (\r{A} GPa$^{-1}$)		&	$6.7047(2), -0.0143(5)$\\
	$c_{0}$ (\r{A}), d$c/$d$P$ (\r{A} GPa$^{-1}$)		&	$9.6809(6), -0.0122(1)$\\
	$\beta_{0}$ (\r{}), d$\beta/$d$P$ (rad GPa$^{-1}$)	&	$113.03(1), -0.35(8)\times 10^{-3}$\\
	$\beta_{11}$ (GPa$^{-1}$)							&	$1.88958\times 10^{-3}$\\
	$\beta_{22}$ (GPa$^{-1}$)							&	$2.13283\times 10^{-3}$\\
	$\beta_{33}$ (GPa$^{-1}$)							&	$1.26021\times 10^{-3}$\\
	$\beta_{13}$ (GPa$^{-1}$)							&	$-0.33984\times 10^{-3}$\\
	$\lambda_{1}$ (GPa$^{-1}$)							&	$2.03806\times 10^{-3}$\\
	$\lambda_{2}$ (GPa$^{-1}$)							&	$2.13283\times 10^{-3}$\\
	$\lambda_{3}$ (GPa$^{-1}$)							&	$1.11173\times 10^{-3}$\\
	$ev(\lambda_{1})$ (GPa$^{-1}$)						&	$(0.98226, \;0, \;-0.42915)$\\
	$ev(\lambda_{2})$ (GPa$^{-1}$)						&	$(0, \;1, \;0)$\\
	$ev(\lambda_{3})$ (GPa$^{-1}$)						&	$(0.42915, \;0, \;0.98227)$\\
	$\psi$ (\r{} from $c$ to $a$)						&	$115.4$\\
	$1/(\beta_{11}+\beta_{22}+\beta_{33})$ (GPa)		&	$189.3$\\
   \hline
   \hline   
  \end{tabular}
  \caption{\label{table:tensor}
   \textbf{
   Isothermal compressibility data for nano-GdBO$\bm{_{3}}$.}
   Ambient pressure unit cell parameters ($x_{0}$)
   and corresponding pressure derivatives (d$x/$d$P$)
   were determined from linear least squares fits to the hydrostatic high-pressure data (shown in Supplementary Fig. \ref{supfig2}).
   Isothermal compressibility tensor coefficients, $\beta_{ii}$, 
   their eigenvalues, $\lambda_{i}$,  
   eigenvectors, $ev(\lambda_{i})$, and 
   angle of direction of maximum compressibility, $\psi$,
   were determined according to the analysis of Ref. \citeonline{Knight2010}.
   }
\end{table}

\newpage
\clearpage

\begin{figure}[b]
\centering
\includegraphics[width=1\textwidth]{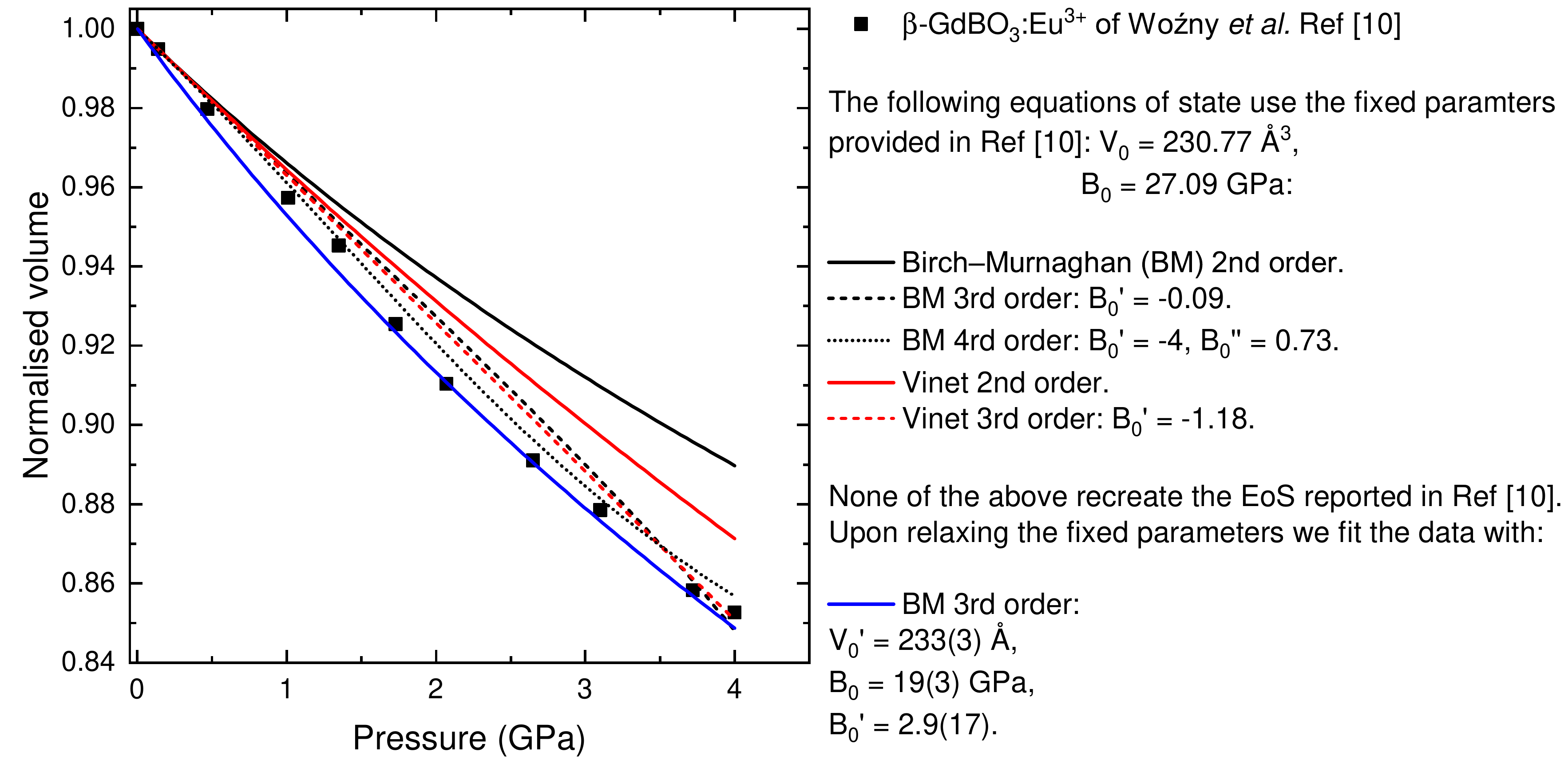}
\caption{\textbf{Different equations of state fitted to the $\bm{\beta}$-GdBO$_{3}$ data of Ref. \citeonline{Wozny2020}.}
 	}\label{fig:wozny_lowBphase}
\end{figure}

\end{document}